# Is the anti-filarial drug diethylcarbamazine useful to treat COVID-19 ?

**Authors:**

**Anuruddha Abeygunasekera**

MBBS (Colombo), MS (Colombo), FRCS (Edinburgh), FRCS (England)

Urological Surgeon, Colombo South Teaching Hospital, Dehiwala 10350, Sri Lanka.

ORCiD number -  0000-0003-3427-6796

E mail: amabey@sltnet.lk

**Saroj Jayasinghe**

MBBS (Colombo), MD (Colombo), FRCP (London), MD (Bristol) PhD (Colombo)

Chair Professor of Medicine, Faculty of Medicine of University of Colombo, Kynsey Road, Colombo 00800, Sri Lanka.

ORCiD number - 0000-0003-1460-6073

Email: saroj@clinmed.cmb.ac.lk

**Corresponding Author:**

Saroj Jayasinghe

Phone: +94-11-2695300

Email: saroj@clinmed.cmb.ac.lk

**Ethical Approval  and Informed Consent:** Not relevant

**Financial support and disclosures:** Personal funds were used to write the paper

**Abstract**

The SARS-CoV-2 virus has resulted in a devastating pandemic of COVID-19. Exploring compounds that could offer a breakthrough in treatment is the need of the hour. Re-positioning cheap, freely available and safe drugs is a priority. The paper proposes evidence for the potential use of diethylcarbamazine (DEC) in the treatment of COVID-19. DEC has inhibitory effects on arachidonic acid metabolism to prostaglandins, little known anti-viral effects on animal retroviruses and demonstrated anti-inflammatory actions in animal models of lung inflammation indicating the need to explore this hypothesis further. We believe this is the first time DEC is being proposed to treat COVID-19.

**Is the anti-filarial drug diethylcarbamazine useful to treat COVID-19 ?**

SARS-CoV-2 virus has caused a pandemic with approximately 414,179 persons affected in 196 countries and 18,440 deaths within a short period of several weeks (1). The virus is a single-stranded RNA virus belonging to the Coronaviridae family, members of which cause mild infections. However, the epidemics caused by Middle East respiratory syndrome coronavirus (MERS-CoV) and severe acute respiratory syndrome coronavirus (SARS-CoV) resulted in alarming morbidity and mortality. COVID-19 caused by SARS-CoV-2 appears to surpass both these in severity. Given the urgency of the outbreak, there is growing interest in repurposing existing agents which have been approved already or to develop novel drugs that can improve the clinical outcome of affected patients.

Key strategies in this regard include development of vaccines to prevent the infection, hot-targeted interventions such as interferon therapies monoclonal antibodies, and small-molecule drugs (2). Despite extensive research being conducted, no antiviral drug had been approved for treating coronavirus MERS-CoV or SARS-CoV and specific interventions for COVID-19 are likely to require several months or even years to be developed (3).

Therefore, repurpose of existing antiviral agents including interferon, chloroquine (an anti-malarial agent) and niclosamide (a broad spectrum anthelminthic) have attracted considerable attention and many trials are underway (4,5). The current paper reviews the evidence to explore the potential of diethyl carbamazine (DEC) to be used successfully as a therapeutic agent for the disease.

**Immune pathogenesis of severe COVID-19**

SARS-CoV-2 has an affinity to the angiotensin converting enzyme 2 (ACE-2) receptors which are expressed in human respiratory epithelia in the lungs. It spreads through the respiratory tract leading to fever, cough, and subsequently in those susceptible to have serious outcomes, it may lead to acute respiratory distress syndrome (ARDS) (6). The pathogenesis of severe disease is considered to be due to the cytokine storm or a dysregulation, in addition to the cytopathic effects of the virus. Both are localized mainly to the lungs due to the presence of high concentrations of virus binding receptors in pneumocytes (7).

A detailed study of individual patients has demonstrated the wide spectrum of the immune response when resolution is associated with recruitment of antibody-secreting cells, T follicular helper cells (TFH) and activated CD4+ and CD8+ T cell populations and elevated Ig M and Ig G SARS-CoV-2-binding antibodies (8).

Many cytokines are implicated in the massive response observed in seriously ill patients. The rapid activation of $CD4^+$ T cells leads to proliferation and differentiation into Th1 cells which secrete proinflammatory cytokines (9). The response consists of high concentrations of IL1B, IFNγ, IP10, and MCP1 (10). Severely ill patients who required intensive care unit (ICU) admission had higher concentrations of granulocyte-macrophage colony-stimulating factor (GCSF), IP10, MCP1, MIP1A, and TNFα than did those not requiring ICU admission. Other studies report the secretion of proinflammatory cytokines such as IL-6, interferon gamma, and granulocyte-macrophage colony-stimulating factor (GM-CSF). GM-CSF activates monocytes to release more IL-6 leading to the formation of a cytokine storm, which triggers ARDS, multi-organ failure (MOF) and even death (9). Furthermore, the infection also initiates an increased secretion of T-helper-2 (Th2) cytokines such as IL4 and IL10. These suppress inflammation and this phenomenon is not seen with SARS-CoV infection (11).

The role of the arachidonic acid related prostaglandin pathways in COVID is less well known. Recent studies have shown age-related changes in this pathway and associated T-cell defects that could account for the increased susceptibility of SARS-CoV infection in the elderly (12). The mechanism involves respiratory dendritic cells (rDC) in the lungs that migrate to the mediastinal lymph nodes and prime T-cells that in turn migrate to the lungs to mount an immune response. An age-related defect in T-cell function is linked to decreased migration of rDC because of increased levels of prostaglandin-D ($PGD_2$) in ageing mice. The resulting poor T cell response is associated with severe infection (13).

**A potential role for diethyl carbamazine (DEC) in COVID-19**

DEC is a cheap and safe drug used for decades in the treatment of filariasis. It is known to have anti-inflammatory actions especially in the lungs, immune-modulatory effects and poorly defined anti-viral effects. The following observations and mechanisms are postulated to consider the role of DEC in treatment of COVID-19.

1. DEC has a wide range of immune-related effects. The main immune modularity mechanism is through its inhibition of lipoxygenase (LOX) and cyclooxygenase (COX) enzymes in the metabolism of arachidonic acid to form prostaglandins including PGD2 (14). Ageing lung is associates with high levels of PGD2, a compound produced from arachidonic acid through the cyclo-oxygenase enzymes. The effects of increased PGD2 include a defect in T-cell function. Since DEC inhibits production of PGD2, it should theoretically enhance T cell responses against respiratory viruses in older humans.

2. DEC therapy has also been shown to enhance antibody production in mice immunized with tetanus toxoid and the cytokine response in animals immunized with LPS both of which facilitates the immune response against microbes (15).

3. Action of DEC on lung injury has been studied using models of acute inflammation. Carrageenan induced pleurisy had increased cellularity, mild haemorrhage and congestion, apoptotic cells, inflammatory cells (mononuclear and polymorphonuclear cells), pulmonary oedema, emphysema and collagen fibers, all of which are attenuated with DEC pre-treatment (16). This is relevant because carrageenan induced pleural inflammation causes an increase in local IL-1 activity in the pleural exudate (17). Its anti-inflammatory effects have been useful in treatment of follicular cystitis of bladder – an inflammatory condition resistant to antibiotics and non-steroidal anti-inflammatory agents (18).

4. In addition, there is some evidence to support the anti-viral activity of DEC against RNA viruses. Mice inoculated with murine leukemic virus, survived significantly longer when they were given DEC (19).

At present there is no proven agent that can eliminate life-threatening pulmonary complications of SARS-CoV-2 infection completely. Several research groups have searched for therapies for infections by SARS-CoV-2 and its complications (2, 4, 20). However, none of these have identified DEC as a potential therapy. This paper argues that it may be worthwhile to consider using DEC as an adjunct to existing drugs to treat COVID-19.